\documentclass[aps,pre,superscriptaddress,10pt]{revtex4-1}
\usepackage[utf8]{inputenc}
\usepackage[english]{babel}

\usepackage{color}
\usepackage{graphicx}
\usepackage{amsmath}
\usepackage{amssymb}
\usepackage{mathrsfs}
\usepackage{fullpage}
\usepackage{relsize}
\usepackage{dcolumn}

\newcommand{\alphaBare}{\alpha_0}
\newcommand{\dBare}{d_0}
\newcommand{\spec}{\mu}  
\newcommand{\numAB}{n_a}
\newcommand{\numPB}{n_p}

\begin{document}

\title{Rectification in a mixture of  active and passive particles subject to a ratchet potential}
\author{Jean-François Derivaux}
\affiliation{DAMTP, Centre for Mathematical Sciences, University of Cambridge,
Wilberforce Road, Cambridge CB3 0WA, United Kingdom}
\author{Robert L. Jack}
\affiliation{DAMTP, Centre for Mathematical Sciences, University of Cambridge,
Wilberforce Road, Cambridge CB3 0WA, United Kingdom}
\affiliation{Yusuf Hamied Department of Chemistry, University of Cambridge, Lensfield
Road, Cambridge CB2 1EW, United Kingdom}
\author{Michael E. Cates}
\affiliation{DAMTP, Centre for Mathematical Sciences, University of Cambridge,
Wilberforce Road, Cambridge CB3 0WA, United Kingdom}

\begin{abstract}
We study by simulation a mixture of active (run-and-tumble) and passive (Brownian) particles with repulsive exclusion interactions in one dimension, subject to a ratchet (smoothed sawtooth) potential.  Such a potential is known to rectify active particles at one-body level, creating a net current in the `easy direction'. This is the direction in which one encounters the lower {\em maximum force} en route to the top of a potential barrier. The exclusion constraint results in single-file motion, so the mean velocities of active and passive particles are identical; we study the effects of activity level, Brownian diffusivity, particle size, initial sequence of active and passive particles, and active/passive concentration ratio on this mean velocity ({\em i.e.}, the current per particle). We show that in some parameter regimes the sign of the current is reversed. This happens when the passive particles are at high temperature and so would cross barriers relatively easily, and without rectification, except that they collide with `cold' active ones, which would otherwise be localized near the potential minima. In this case, the reversed current arises because hot passive particles push cold active ones preferentially in the direction with the lower spatial separation between the bottom and top of the barrier. A qualitatively similar mechanism operates in a mixture containing passive particles of two very different temperatures, although there is no quantitative mapping between that case and the systems studied here. 
\end{abstract}
\maketitle

\section{Introduction}

Recent years have seen an important surge in the study of active systems, which encompass a broad variety of natural and artificial cases, including but not limited to bird flocks, fish schools, motile bacteria, or self-propelled synthetic colloids.
While these systems span different length scales, they share a common property: their individual constituents are out of equilibrium locally, dissipating energy to generate motion. This feature results in a new and rich phenomenology: collective oriented motion \cite{wang_spontaneous_2011}, motility-induced phase separation \cite{bialke_microscopic_2013,cates_motility-induced_2015}, active turbulence \cite{wensink_meso-scale_2012} and chemotaxis \cite{liebchen_synthetic_2018}, to cite only a few. 

Active particles can also strongly impact the behaviour of any passive particles with which they are mixed \cite{bechinger_active_2016}. This influence already shows up in the enhanced diffusion of single passive particles in an active bath \cite{leptos_dynamics_2009,zhao_enhanced_2017}, and in sustaining the rotation of a passive microgear in a bacterial suspension \cite{leonardo_bacterial_2010}. At higher densities of the passive particles, phase separation of the binary mixture occurs at large P\'eclet number, giving rise to dynamical clusters of passive particles surrounded by an active corona \cite{wysocki_propagating_2016,stenhammar_activity-induced_2015,wittkowski_nonequilibrium_2017}. These clusters depart from the more traditional phase separation of purely passive particles, as merging and melting happen unremittingly at the interfaces. Indeed, dynamical instability is a distinctive feature of active-passive mixtures, whose demixing admits travelling-wave modes,which are forbidden in a binary passive mixture \cite{you_nonreciprocity_2020,saha_scalar_2020}. Hence, active particles are interesting candidates for modifying transport or enhancing separation of passive particles on the microscale \cite{wang_review_2019,banerjee_tracer_2022}. Conversely, passive particles can be used to modulate and control the effects of activity.

\begin{figure}
\includegraphics[width = 0.5 \textwidth]{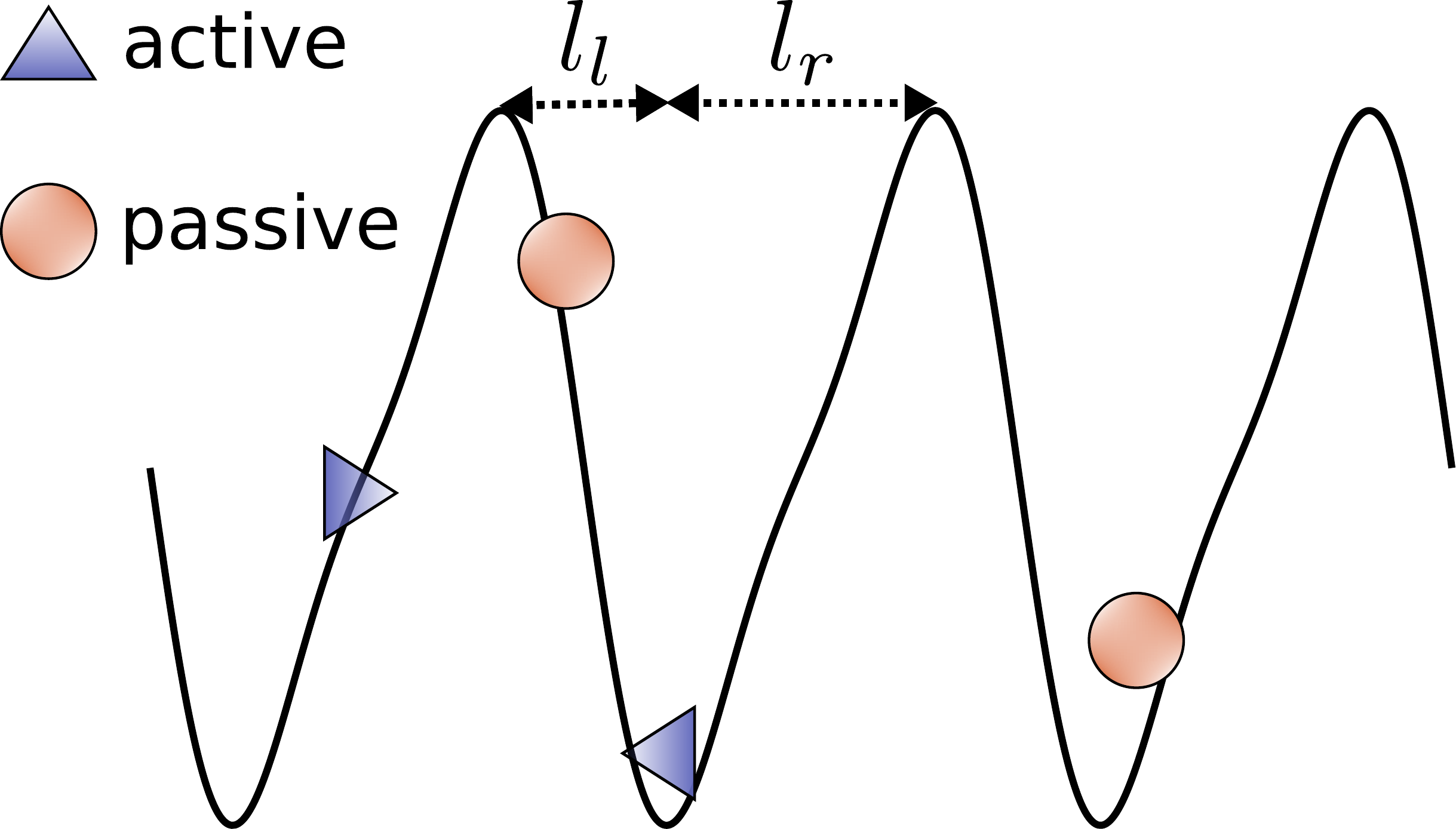}
\caption{Schematic of the smoothed sawtooth ratchet potential studied in this paper. A mixture of active (triangles) and passive (circles) particles move in an asymmetric potential $V(x) = - \frac{V_0}{2\pi} \left[ \text{sin}(2\pi kx) + \frac{1}{4}\text{sin}(4\pi kx)\right]$. Note that $l_r = 0.62 \; k^{-1}$ and $l_l= 0.38 \; k^{-1}$.}
\label{fig:sawtooth_potential}
\end{figure}

In this paper, we will study the transport properties of an active-passive mixture subjected to a ratchet potential, comprising a smooth sawtooth shape in one dimension (Fig.~\ref{fig:sawtooth_potential}). It is now established that active particles alone can generate a non-zero current in such a potential, even without interactions among the particles \cite{reichhardt_ratchet_2017}. Numerical and experimental evidences of active rectification were found in a variety of ratchet geometries, such as along curved walls \cite{nikola_active_2016}, in pores \cite{ghosh_self-propelled_2013} or funnels \cite{galajda_wall_2007}, as well as in sawtooth-shaped potentials \cite{mcdermott_collective_2016}. Separately, previous studies showed that a mixture of two species of passive particle, with one sensitive to a ratchet and the other not, could lead to non-trivial currents in the system, depending on the interactions between particles \cite{savelev_controlling_2003,savelev_manipulating_2004,savelev_controlling_2005,marchesoni_driven_2006,hanggi_artificial_2009}. 
Similar effects can also be observed in systems of anisotropically pinned magnetic flux quanta in superconductors, subjected to a.c. fields \cite{villegas_superconducting_2003,de_souza_silva_controlled_2006}.   These systems support effective ratchet potentials that can induce vortex motion in either direction, depending on the concentration of the vortices and the field strength.

However, less is known about the rectification of active-passive mixtures. Due to their out-of-equilibrium intrinsic nature, active particles should offer new pathways to rectify passive particles. Non-zero edge currents can be generated in a 2D mixture of chiral active particles and Brownian disks, by tuning the concentrations or imposing a temperature gradient \cite{reichhardt_reversibility_2019,zhu_rectification_2020}. Use of external, asymmetric potentials is also promising to sort mixtures; indeed, in recent experiments bacteria were shown to  assist the transport of passive colloidal beads towards asymmetric micrometric obstacles \cite{koumakis_targeted_2013}. Less sharp but asymmetric obstacles, such as half-disks, can also rectify passive particles in a mixture \cite{rojas-vega_fast_2021}. A better understanding of the parameters influencing the passive particle flux induced by active motion is then necessary to control more accurately the currents in the system.

We therefore study here a minimal, 1D model of an active-passive mixture, with strongly repulsive interactions (that prevent particles crossing), in a smoothed sawtooth potential. The active particles are run-and-tumble particles (RTPs) and the passive ones are Brownian. Compared to previous studies on rectification in mixtures of interacting passive particles, involving complex external potentials, the model here is simpler, and does not require an external drive. (Instead, the non-equilibrium forces come directly from the active particles.)

In Section~\ref{sec:model}, we specify the model and express it in non-dimensional units. We also give a precise definition of the (averaged) currents that are our subject of study.  In Section~\ref{sec:results}, we use extensive numerical simulations to explore how the different parameters of the system (tumbling rate, diffusion coefficients, particles sizes, ...) influence those currents. Our goal is not a complete survey of the high-dimensional parameter space but rather to identify certain regimes in which the currents are large enough to be unambiguously identified from the numerical data, and then to gain mechanistic insight by parameter variation across such regimes. 
An important feature of the active-passive mixture is that the current may flow in either direction; this contrasts with the situation for active particles alone, where particles always travel down the steeper side of the potential, and up the shallower one.  The unexpected current-reversal in the mixture can be explained on the basis of temperature differences between active and passive particles. In these regimes, 'cold' active particles are trapped into potential minima, and have more probability to be pushed to the left than to right because of the shorter separation with the potential maxima. Given the importance of temperature difference in this phenomenon, we draw a comparison with a mixture of passive particles at two different temperatures that also shows negative currents by a qualitatively similar mechanism.
Conclusions and perspectives are presented in Section~\ref{sec:conclusions}.


\section{Model}
\label{sec:model}

We study a mixture of active and passive particles subjected to a ratchet potential in a 1D box of length $L$, with periodic boundary conditions. 
A schematic of the system is depicted in Fig.~\ref{fig:sawtooth_potential}. 
The number of active particles (RTPs) is $N_a$ and the number of passive (Brownian) particles is $N_p$.  
In addition to their random (active or passive) motion, all particles feel a ratchet force $F_{rat}$ (dependent on the particle position), and interact through repulsive forces $F_{int}$, which depend on their positions relative to the other particles.  The equations of motion for active and passive particles positions are 
\begin{align}
\dot{x}_{a,i} & = \frac{1}{\gamma_a} \left[ F_{rat}({x}_{a,i}) + F_{int,i} \right] + v_0  \, \sigma_i (t) \label{eq:evo_active}  \\
\dot{x}_{p,j} & = \frac{1}{\gamma_p} \left[ F_{rat}({x}_{p,j}) + F_{int,j} \right] + \sqrt{2D_p} \, \xi_j (t) \;. \label{eq:evo_passive}
\end{align}
where $x_{a,i}$ is the position of the $i$th active particle and similarly $x_{p,j}$ for the $j$th passive particle; also $\gamma_a$ and $\gamma_p$ are the friction coefficients the two types of particle, $v_0$ is the bare self-propulsion speed of active particles, and $D_p$ is the bare diffusivity of passive particles. The stochastic forces $\sigma,\xi$ depend on the nature of the particles and are specified below, as are the details of the ratchet and interaction forces.

The ratchet force is the gradient of a differentiable sawtooth potential 
\begin{equation}
V(x) = - \frac{V_0}{2\pi} \left[ \text{sin}(2\pi kx) + \frac{1}{4}\text{sin}(4\pi kx)\right] \; ,
\label{eq:sawtooth_pot}
\end{equation}
so that
\begin{equation}
F_{rat}(x) = - \nabla V(x) =  k V_0   \left[ \text{cos}\left( 2 \pi k x \right) + \frac{1}{2} \text{cos}\left( 4 \pi k x\right) \right] \; .
\end{equation}
The interaction forces are modelled by
a repulsive Weeks-Chandler-Andersen (WCA) potential of strength $\epsilon$: 
\begin{equation}
V_{\text{WCA}}\!\left( r \right) = 4\epsilon \left[ 
\left( \frac{\dBare}{r}\right)^{12} -  \left( \frac{\dBare}{r}\right)^6 + \frac14  \right] \theta\!\left(2^{1/6}\dBare - r \right)
\label{eq:WCA_potential}
\end{equation}
where $r$ is the distance between two particles, $\dBare$ is the particle diameter (which is the same for all particles) and $\theta$ is the Heaviside function. 
Then the interaction force on particle $i$ is
\begin{equation}
F_{int,i} = \frac{48 \epsilon}{\dBare} \mathlarger{\sum}_{l (\neq i)}  \left[ \left( \frac{\dBare}{r_{il}} \right)^{13} - \frac12 \left(\frac{\dBare}{r_{il}} \right)^{7} \right] \theta\!\left(2^{1/6}\dBare - \vert r_{il} \vert \right)
\end{equation}
where the sum runs over all particles in the system except particle $i$ itself (this includes both active and passive particles), and $r_{il}=x_l-x_i$.  

The stochastic forces are different for active and passive particles.  In \eqref{eq:evo_active}, $\sigma_i$ indicates the orientation of the RTP, whose possible values are $\pm1$.  The orientation changes sign at random with rate $\alphaBare$, which is the tumbling rate.  Each particle tumbles independently so the covariance of the orientation at different times decays exponentially:
\begin{equation}
\langle \sigma_i(t) \sigma_j(t^{\prime}) \rangle = e^{-2\alphaBare \vert t-t^{\prime}\vert} \delta_{ij} \;.
\end{equation}
For free RTPs, this leads to a diffusivity at long times equal to
\begin{equation}
D_a = \frac{v_0^2}{2\alphaBare} \; .
\label{equ:Da}
\end{equation}

By contrast, the $\xi_j$ in \eqref{eq:evo_passive} are independent Gaussian white noises with
\begin{equation}
\langle \xi_i(t) \xi_j(t^{\prime}) \rangle = \delta_{ij} \delta(t-t^{\prime}) \, .
\end{equation}
so that passive particles have diffusivity $D_p$.  Under the assumption that this stochastic force originates in thermal fluctuations, we identify the temperature of the corresponding heat bath as
\begin{equation}
T = \frac{\gamma_p D_p}{k_B} \; .
\label{equ:T0}
\end{equation} 

Note that the active particles have no thermal diffusion, within this model.  Thus in the limit of low activity, $D_a\to 0$, we approach a mixture of two passive species, one of which is held at temperature $T$ and the other at zero temperature.  This limit can be reached either by taking the propulsion speed $v_0\to0$ or the tumbling rate $\alphaBare\to\infty$ (with other parameters held constant).  In the following we fix $v_0$ so this passive limit is $\alphaBare\to\infty$.

\subsection{Nondimensionalised model}

The parameter space of the model can be simplified by introducing non-dimensional space and time variables $x^{\star} = k x$ and $t^{\star} = k v_0 t$. Lengths are thus rescaled by the wavelength of the potential, and times by the time it takes a free active particle to translate one such wavelength.
Rescaling the forces by the self-propulsion force $\gamma_a v_0$ yields
\begin{align}
F_{rat}^{\star}(x^\star) & = f \left[ \text{cos}\left( 2 \pi x^{\star}\right) + \frac{1}{2} \text{cos}\left( 4 \pi x^{\star}\right) \right] \\
F_{int,i}^{\star} & = \epsilon_0 d \, \mathlarger{\sum}_{l (\neq i)} \left[ \left( \frac{d}{r^{\star}_{il}} \right)^{13} - \frac{1}{2}  \left( \frac{d}{r^{\star}_{il}} \right)^{7} \right] \theta\!\left(2^{1/6}d - \vert r^\star_{il}\right\vert) \, ,
\end{align}
in which we introduced the following rescaled parameters:
\begin{equation}
f = \frac{V_0 k}{\gamma_a v_0}\,, \qquad d = k\dBare\,, \qquad  \epsilon_0 = \frac{48 \epsilon }{\gamma_a v_0 k \left( \dBare\right)^{2} }\, .
\end{equation}
In those new rescaled forces, $f$ is the amplitude of the ratchet force,  $d$ is the particle diameter, and $\epsilon_0$ the interaction force amplitude.
Then Eqs.~(\ref{eq:evo_active}-\ref{eq:evo_passive}) become
\begin{align}
\frac{dx^{\star}_{a,i}}{dt^{\star}} & = F_{rat}^{\star}(x^{\star}_{a,i}) +  F_{int,i}^{\star} + \sigma_i(t^\star) \label{eq:evo_active_nondim}\\
\frac{dx^{\star}_{p,j}}{dt^{\star}} & = \Gamma \left[ F_{rat}^{\star}(x^{\star}_{p,j})  +  F_{int,j}^{\star} \right]
 + \ell_{D} \, \xi_j^\star(t^\star) \label{eq:evo_passive_nondim} 
\end{align}
where the rescaled noise obeys $\langle \xi_i^\star(t^\star) \xi_j^\star({t^\star}') = \delta_{ij} \delta(t^\star-{t^{\star}}')$,
and we introduced further non-dimensional parameters:
\begin{equation}
\label{eq:nondim_params}
\Gamma = \frac{\gamma_a}{\gamma_p}\,, \qquad \alpha = \frac{\alphaBare}{k v_0}\,,  \qquad \ell_D = \sqrt{\frac{2D_pk}{v_0}} \;.
\end{equation}
Here $\Gamma$ is the ratio of mobilities of passive and active particles, $\alpha$ is the tumbling rate, and $\ell_D$ is the noise amplitude (all in rescaled units). The last of these is the root-mean-square diffusive displacement during the time taken by a free active particle to move one wavelength without tumbling. Thus the (rescaled) distance $\ell_D$ is a measure of the thermal agitation of passive particles.

In the rest of this paper, we will only refer to the nondimensionalised model and its parameters. Consequently, we will drop the star notation from here on.
To full specify a system of $N=N_a+N_p$ particles, we define also
\begin{equation}
\phi_a = \frac{N_a}{N}, \qquad c = \frac{N}{k L}
\end{equation} 
which are the fraction of particles that are active, and the (dimensionless) concentration, respectively.
The full system is then characterised by 8 non-dimensional parameters $ \left(\Gamma,  f, \alpha, \epsilon_{0}, d, \ell_D , \phi_a, c \right)$. 

Since this resulting parameter space is too large for systematic exploration, we focus on a restricted (but representative) subspace.  We consider active and passive particles with the same mobility $\Gamma=1$, and we also fix $c=1$ (one particle per wavelength of the potential).  We mostly focus on the average particle current in the system (see below), and its dependence on $f$, $\alpha$, $\ell_D$ and $d$.  We briefly explore the roles of $\phi_a$ and $\epsilon_0$ in Appendix~\ref{app}.

\subsection{Measuring the current}

The central quantity in our study of this system is the steady-state average particle velocity, or current per particle, $\langle v\rangle$. Note that in principle one can define separate active and passive currents $\langle v_{a,p} \rangle $, but since our system has periodic boundary conditions, and since the repulsive interactions (which diverge at $r=0$) prevent any particle from crossing trajectories with its neighbours, the time-averaged active and passive currents must be identical so we use the same symbol $\langle v\rangle$ for both. 

Nonetheless, there are various different ways to extract the average current from a simulation. First one may compute the the time average of the speed of particle $i$:
\begin{equation}
\overline{v}_{\spec,i} = \frac{1}{t} \int_{0}^{t} v_{\spec,i}(t^{\prime}) \, dt^{\prime} \; ,
\end{equation}
where $\spec \in (a,p)$. In a second step, we may take the average of $\overline{v}_{\spec,i}$ over the set of particles of the same species, leading to the particle-averaged current,
$
\overline{v}_{\spec} = \frac{1}{N_{\spec}} \sum_{i}^{N_{\spec}} \overline v_{\spec,i} \; .
$

For a simulation run of sufficient duration the two quantities $\overline{v}_{\spec,i}$ and $\overline{v}_{\spec}$ coincide, and furthermore do not depend on the index $\spec$. However, they can depend on the sequence of active and passive particles in the system, which is the only part of the initial condition conserved by the dynamics. We explore this later, and show that this sequence-dependence is weak in most regimes of interest. Accordingly we define $\langle v\rangle  = \langle \overline v_{\spec} \rangle$ as the ensemble average over runs with randomly chosen sequences of active and passive particles, of the time- and particle- averaged current $\overline{v}_{\spec} $. This additional averaging allows numerical errors to be reduced so that clearer trends in the  current are discernible. Despite this, $\langle v \rangle$ is a somewhat noisy quantity for many of our parameter sets. Our focus is on cases where the mean current is large enough in magnitude that we can at least be confident of its sign.

We have carried out extensive numerical simulations of equations (\ref{eq:evo_active_nondim}-\ref{eq:evo_passive_nondim}), integrated with the Euler-Maruyama scheme. For each parameter set, the current was averaged over a definite number of runs $N_{\text{runs}}$. In each run, initial orientations of active particles and initial sequence of active and passive particles were different and sampled from a uniform distribution. The initial spatial distribution of particles, whatever their nature, was set to a homogeneous distribution: each particle is spaced from the others by the same distance. In a run, we then allowed the system to relax to a steady state during an equilibration time $t_{\text{eq}}$. After this time, the trajectory was recorded to compute the time-averaged current of this run during a time interval $t$. A final averaging was carried out over all $N_{runs}$ time evolutions to obtain the final average $\langle v \rangle$. Error bars, where depicted in the following figures, correspond to an interval of $\pm$ the standard error of the mean (SEM) of the run-wise averaged data.

\section{Results}
\label{sec:results}

\subsection{Role of diffusivities: Positive and negative currents}
\label{APM}

As mentioned in the Introduction, it is well-known that noninteracting active particles can be rectified in a ratchet potential to give a non-zero current \cite{reichhardt_ratchet_2017,mcdermott_collective_2016,ghosh_self-propelled_2013}. For the specific ratchet potential \eqref{eq:sawtooth_pot}, strong rectification was reported for 2D active Brownian particles (ABPs) within a certain range of intermediate values of the ratchet force $f$ \cite{mcdermott_collective_2016}. If $f$ is too small the current becomes insignificant and if $f$ is too large it vanishes identically. 

\begin{figure}
\includegraphics[width = 0.9 \textwidth]{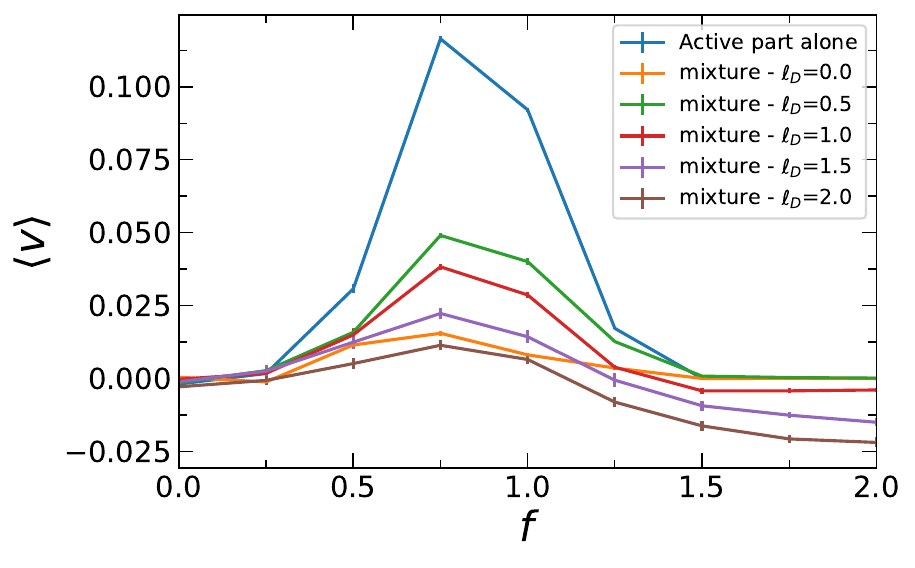}
\caption{Dependence of the current $\langle v\rangle$ on $f$. The current strongly depends on the passive diffusivity via the parameter $\ell_D$. 
Parameters: $\Gamma=1$, $\alpha=  1 $, $\epsilon_{0} = 0.25$, $d = 0.2$, $N_a = N_p = 10$, $\phi_{a} = 0.5$, $c=1$, $t = 400$, $t_{\text{eq}} = 100$, $dt = 5 \times 10^{-6}$, $N_{runs} = 40$. }
\label{fig:vP_alpha1}
\end{figure}

Fig.~\ref{fig:vP_alpha1} demonstrates the same effect in simulations of 1D RTPs alone, corresponding to $\phi_a=1$.
The mechanism is the same as the RTP and ABP cases. When the ratchet force is weak, active particles are insensitive to the potential and undergo almost free diffusion. Nonetheless their trajectories do not obey detailed balance so that a weak periodic potential without inversion symmetry will generically lead to a finite rectification current \cite{julicher_modeling_1997}. For this to become large, the mean speed of left and right moving particles have to be significantly different, which requires $f$ of order unity in our units. Quantitatively, for noninteracting particles $\langle v\rangle$ reaches its maximum value at $f = 2/3$, where the maximal force exerted on the particle by the left (steep) side of the potential equals the self-propulsion speed $v_{0}$. Particles are then unable to exit a well by moving left but can still do so by moving to the right. In the following, we refer to the $+x$ (right) direction as the `easy' direction, because a particle moving away from the potential minimum feels a smaller force if it travels in this direction.  Similarly, the $-x$ (left) direction is referred to as the `difficult' direction.

On increasing $f>2/3$, the ratchet force opposing rightward motion grows, so the current decreases until $f=4/3$, where escape to the right also becomes impossible (in a non-interacting system). For larger $f$, all particles are localized and the current is zero. The sharpness of this transition is associated with the fact that active particles have a fixed (or fixed maximum) speed; it is smoothed out by any additional Brownian component to their motion, not present in our model. (Interactions can also have this effect in some regimes.) Under the conditions just described, the current is maximal when tumbling is rare enough that a particle can traverse a full wavelength of the potential without reversing. When instead the persistence length $\alpha^{-1}$ of the (free) active particle motion is much smaller than one wavelength, particles spend most of the time in quasi-diffusive exploration around the potential minima with only occasional excursions up the potential during an unusually long tumble-free interval. 

The above scenario for active-only systems is clearly perturbed in a system where some of the particles are passive. For example, with hardcore interactions between the two species, the  Brownian motion of a passive neighbour can push an active particle over a barrier that it cannot otherwise cross. Conversely, collisions with active particles can impart asymmetric motion to their passive (Brownian) neighbours which, without such collisions, would be in thermal equilibrium and therefore show no steady-state currents in any periodic potential.
In the mixed system we expect a role to be played by $\ell_D$, the averaged distance undergone by a passive particle during the time taken for a free active particle to cross one wavelength. This nondimensionalized `thermal length' characterizes the relative strength of the passive Brownian motion, rather as the persistence length $\alpha^{-1}$ characterizes activity. However, since the active and passive particles respond to the ratchet potential in quite different ways, the direct comparison of these two lengths is not always a useful guide to mechanism.

We first investigate the case $\alpha=1$, and we vary the passive diffusivity by changing $\ell_D$ in an equal mixture of active and passive particles ($\phi_a = 0.5$). The interaction parameters are held fixed, with a hard-core particle diameter $d=0.2$ (one fifth of the lengthscale of the ratchet). 
In Fig. \ref{fig:vP_alpha1}, we plot the dependence of the current per particle $\langle v\rangle$ on the ratchet force $f$, for various $\ell_D$ in the range $[0,2]$.
The diffusivity of the passive particles strongly influences the resulting curves, in an interesting and nonmonotonic way. First, when the passive particles do not diffuse, the current follows broadly the same trend as for active particles alone: a net current to the right is generated for intermediate values of $f$, which vanishes (within numerical error) for $f\gtrsim 1.5$. The magnitude of this current is however greatly reduced because the passive particles have to be pushed across the barriers due to the hardcore constraints, obstructing the active dynamics, with the passive particles then rectified alongside the active ones.
When the passive diffusivity is moderately increased ($\ell_D = 0.5$), the current initially increases, presumably because their thermal motion makes the passive particles easier to push around. However, raising the diffusivity further, we find that the current starts to fall, not only for intermediate $f$ where it stays positive, but also for large $f$ where a negative current is clearly apparent for $\ell_D\gtrsim1.0$.

\begin{figure}
\includegraphics[width = 0.9 \textwidth]{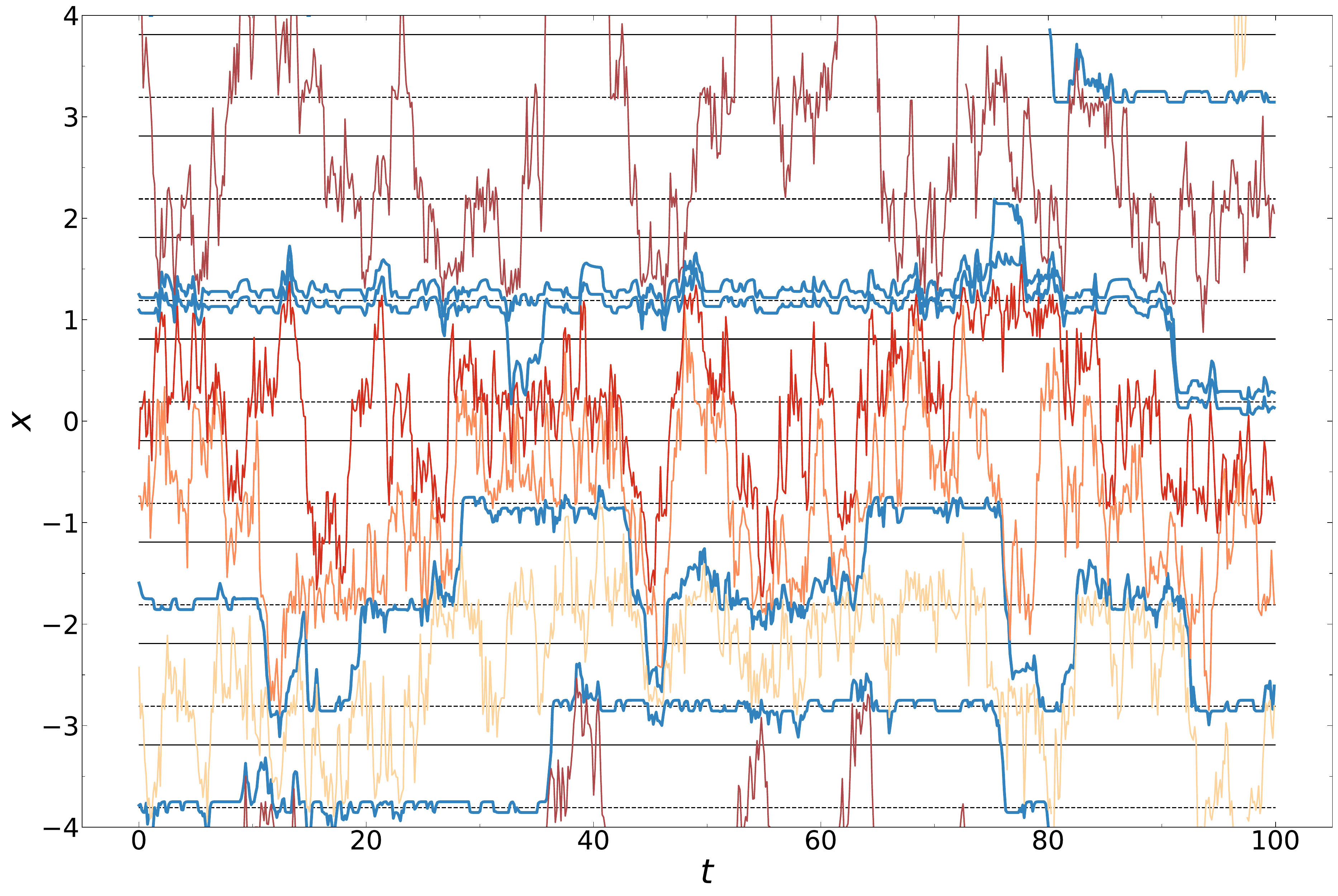}
\caption{Time series of 4 active particles (blue trajectories) and 4 passive particles (shades of orange and red) at $\ell_D=1$, where passive rectification dominates. Dotted lines stand for the potential minima and the plain lines for the potential maxima. Due to the large fluctuations of passive particles, the general movement is oriented towards negative positions.
Parameters: $f =2$, $\ell_D=1$, $\Gamma=1$, $\epsilon_{0} = 0.25$, $d = 0.2$, $N_a = N_p = 4$, $\phi_{a} = 0.5$, $c=1$, $t = 100$, $t_{\text{eq}} = 20$, $dt = 5 \times 10^{-6}$. }
\label{fig:seriet_A1.50}
\end{figure}

\subsection{Mechanism of negative current: `cold' active particles are pushed in the `difficult' direction}

Negative currents were already observed for 2D ABPs \cite{mcdermott_collective_2016} and for single-file  Brownian particles subject to time-periodic forcing \cite{derenyi_cooperative_1995}, both in a similar potential. These effects occur primarily at high concentration.  In our system, the effective volume fraction is $c d=0.2$, so the negative current cannot be attributed to crowding effects. 

Instead, 
the reversal of the current can be explained by the following argument.  First, recall from (\ref{equ:T0}) that increasing $\ell_D$ (and hence $D_p$)
represents an increase in the temperature of the passive particles.  This makes them increasingly insensitive to the ratchet potential: without the active particles the passive ones would move right or left with (equal) high probability. Meanwhile, because $f$ is also large, each active particle is trapped in a potential well until a `hot' passive particle pushes it across one of the barriers.  
Because the spatial distance $l_l$ from the bottom of a well to the top of the barrier to its left is shorter than the distance $l_r$ to the right one (see Fig.~\ref{fig:sawtooth_potential}), active particles move leftward on average, despite this being the `difficult' direction for the active dynamics.

Fig.~\ref{fig:seriet_A1.50} illustrates this effect in a system of 4 active and 4 passive particles in a case where the ratchet force and the passive diffusion coefficient are both high ($f = 2$ and $\ell_D=1$). The current is clearly in the `difficult' ($-x$) direction, with the passive particles chasing active particles from the potential wells (dotted lines). At the end of the trajectory, each particle has moved to the next well to the left when compared to the initial position.

In this situation, it is natural to draw an analogy with thermal systems: since the active particles are trapped in potential minima, they behave similarly to passive particles with a low temperature.  Hence we refer to them as `cold' active particles.  (Despite this suggestive nomenclature, it is important to keep in mind that the particles are not thermal, for example their fluctuations near the potential minima differ strongly from Boltzmann statistics.)

\subsection{Active and passive rectification}

\begin{figure}
\includegraphics[width = 1.0 \textwidth]{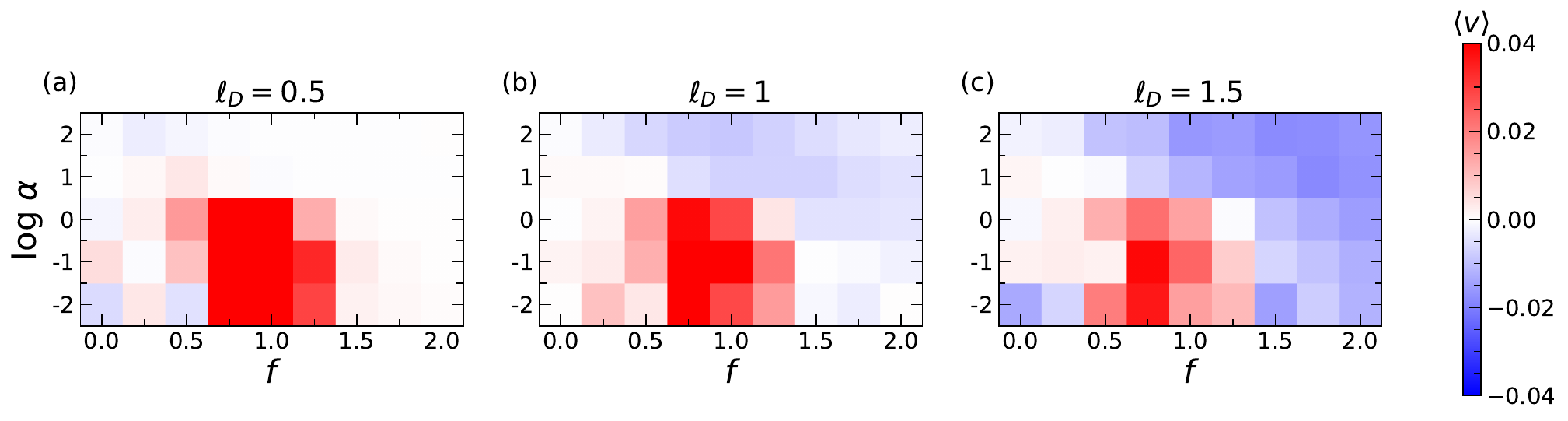}
\caption{Maps of the current depending on the ratchet force $f$ and on the tumbling rate $\alpha$. (a) $\ell_D = 0.5$, (b) $\ell_D=1$ and (c) $\ell_D=1.5$.
Parameters: $\Gamma=1$, $\epsilon_{0} = 0.25$, $d = 0.2$, $N_a = N_p = 10$, $\phi_{a} = 0.5$, $c=1$, $t = 400$, $t_{\text{eq}} = 400$, $dt = 5 \times 10^{-6}$. 
}\label{fig:PD_alphaP}
\end{figure}

Given that this negative current takes place because active particles are pushed across the barrier by passive ones, one may expect a similar effect in mixtures of hot and cold passive particles.  Such an effect is demonstrated in Section~\ref{2temp} below.  For this reason,
we call this regime at large $f$ and large $\ell_D$ the `passive rectification' regime, in contrast to the `active rectification' seen in the forward direction at moderate $f$ and small $\ell_D$.  For any given $f$, active rectification operates most effectively when $\alpha$ is small so that the active motion is highly persistent. Conversely, large $\alpha$ helps to localize particles near the potential minima \cite{cates_motility-induced_2015}. Large $\alpha$ and $\ell_D$ represent the optimal conditions for passive rectification, which then can occur across the full range of $f$. This is shown in the diagrams of  Figure~\ref{fig:PD_alphaP} which map out the sign and magnitude of the current in the $(f,\log\alpha)$ plane for various $\ell_D$. Whereas at low passive temperature (Fig.~\ref{fig:PD_alphaP}a), the currents are either positive (intermediate $f$ and small $\alpha$) or insignificant (elsewhere), the domain and strength of the negative currents increases with $\ell_D$ (Fig.~\ref{fig:PD_alphaP}b,c).

These results show that for any given $f$, the switch from active to passive rectification mode is governed by 
$\alpha \propto D_a^{-1}$ and $\ell_D \propto D_p^{1/2}$, where $D_{a,p}$ are the diffusivities of (free) active and passive particles [recall (\ref{equ:Da},\ref{eq:nondim_params})]. One obvious idea for reducing the dimensions of the parameter space is to test whether there is data collapse via the combined parameter
$\mathcal{D} = {D_a}/{D_p} = 1/(\alpha \, \ell_D^{2})$.
This parameter was used to analyze demixing of particles at different temperature  in previous studies \cite{weber_binary_2016-1,tanaka_hot_2017}. 

\begin{figure}
\includegraphics[width = 0.9 \textwidth]{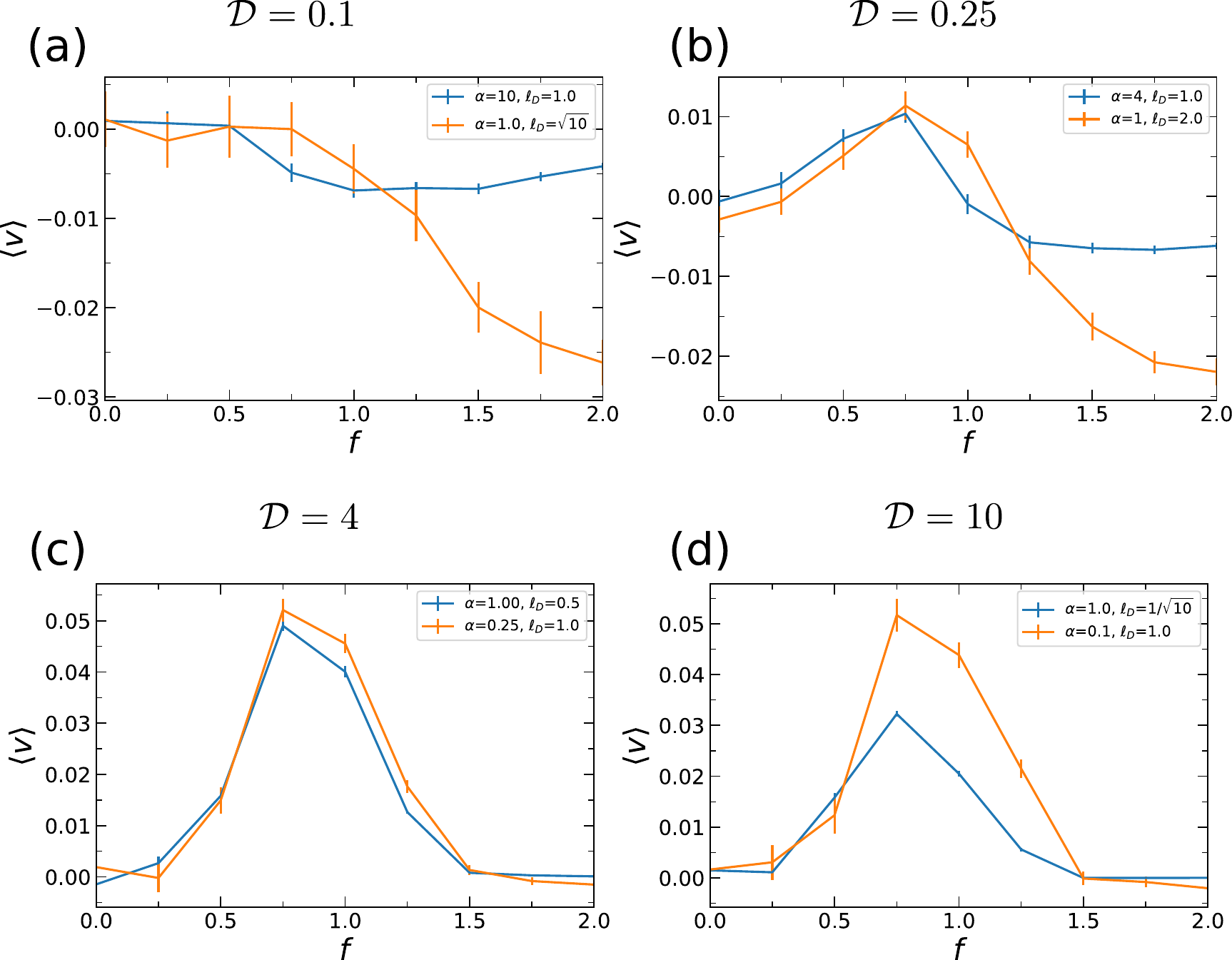}
\caption{ Current versus $f$ curves when the active and passive bare diffusion coefficients are varied while holding their ratio constant. The ratio $\mathcal{D}$ alone is not sufficient to describe the data.
Parameters: $\Gamma=1$, $\epsilon_{0} = 0.25$, $d = 0.2$, $N_a = N_p = 10$, $\phi_{a} = 0.5$, $c = 1$, $t = 400$, $t_{\text{eq}} = 100$, $N_{runs} = 40$. All simulations used a timestep $dt=5 \times 10^{-6}$ except for the  oranges curves in (a) ($dt=1 \times 10^{-6}$), and in (b) ($dt=2.5 \times 10^{-6}$).}
\label{fig:ratio_diff}
\end{figure}

In Fig.~\ref{fig:ratio_diff} we show current versus $f$ curves for two datasets at each of various $\mathcal{D}$ and find that the data collapse fails for both small and large $\mathcal{D}$ values.
 For the latter, Fig.~\ref{fig:ratio_diff}(d), the active rectification is strong and large currents can be maintained as long as the ratchet force to the right side of a minimum does not reach the stalling force (in our units, the self-propulsion speed). Here, if the tumbling rate $\alpha$ is raised (so that particles are more likely to reverse before crossing a barrier), and $D_p$ reduced so as to keep $\mathcal{D}$ the same, the $D_p$ reduction disfavours passive rectification but this does not compensate the diminished active rectification effect and the total current decreases. Likewise for $\mathcal{D}$ less than one, Fig.~\ref{fig:ratio_diff}(a,b), the passive rectification current is an increasing function of $D_p$ that cannot be cancelled by a matched increase in $D_a$. 
 
The generation of current by tuning diffusion coefficient was envisioned by Büttiker several decades ago \cite{buttiker_transport_1987}. In that work, a spatially modulated diffusion coefficient for a single species in a symmetric potential can induce an effective asymmetric potential, leading to a current. However, this situation differs from the active-passive mixture, in which particles have different but spatially constant bare diffusivities. As highlighted above, the current mechanism relies primarily on a kinetic effect, related to interaction between the two different species, which is qualitatively different from that of~\cite{buttiker_transport_1987}.
 
In summary, the rectification regimes rely on mechanisms that are more subtle than a simple diffusion difference between two different species. This is why there is no data collapse via $\mathcal{D}$ and the dependences on $D_a$ and $D_p$ (or equivalently, $\alpha$ and $\ell_D$) must be considered separately, as was done above. The inability to capture the physics behind the rectification with diffusion differences will be further investigated in Section \ref{2temp} by directly comparing the active-passive mixture with a passive-passive mixture.

\subsection{Role of particle size}

In our system, active and passive particles share the same diameter $d$. As mentioned earlier, this non-dimensional size is the ratio between the particle size and the wavelength of the ratchet potential. (Size has an important influence on the escape rate from a confining potential for ABPs through the control of rotational diffusion \cite{geiseler_kramers_2016}. Larger particles have longer persistence length and escape more easily. Such an effect is obviously not expected in a 1D system). Particle size does not affect the ratchet force on a particle which depends only on the location of its centre. It does however affect the collisions between particles, and hence the ability for one particle to push another across a barrier. (Notably, for large enough $d$ the particle doing the pushing need not be near the top of a barrier itself.) In this Section we use numerical simulations to assess this size effect.

\begin{figure}
\includegraphics[width = 0.9 \textwidth]{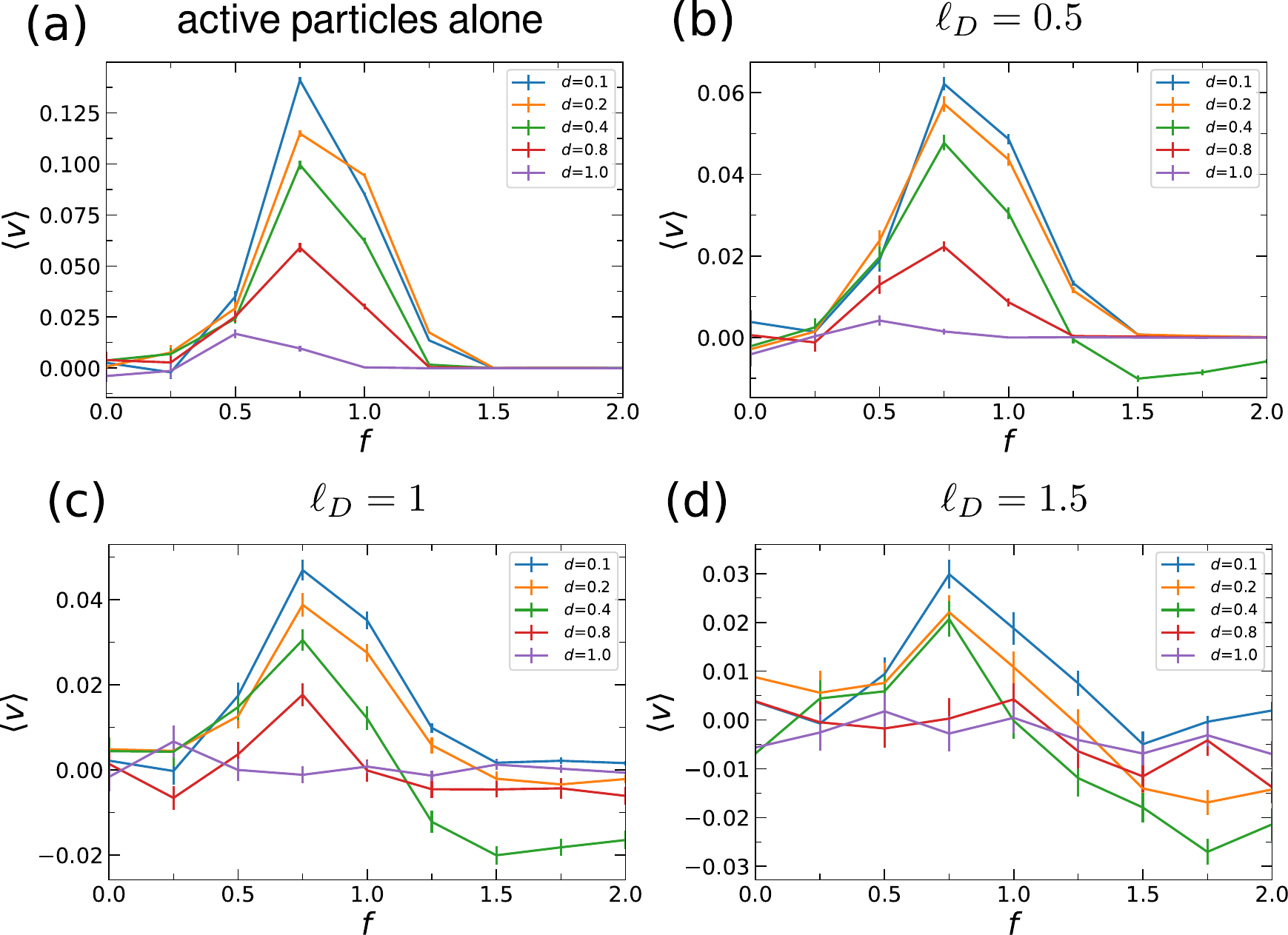}
\caption{Effect of particles size on the currents (a) for active particles alone, (b) for weak passive diffusion ($\ell_D = 0.5$), (c) for medium passive diffusion ($\ell_D = 1$) and (d) for high passive diffusion ($\ell_D = 1.5$).
Parameters: $\Gamma=1$, $\epsilon_{0} = 0.25$, $N_a = N_p = 10$, $\phi_{a} = 0.5$, $c=1$, $t = 100$, $t_{\text{eq}} = 20$,  $dt = 1 \times 10^{-6}$, $N_{runs} = 40$. }
\label{fig:size}
\end{figure}

Our results are displayed in Fig.~\ref{fig:size}. For a system composed of active particles alone (Fig.~\ref{fig:size}a), the change of size impacts the magnitude of the active rectification current without changing its direction. Repulsions exerted by larger particles across larger distances hinder the motion and tend to enhance trapping effects. This picture changes dramatically in an active/passive mixture. 
In the regime where active rectification operates (Fig.~\ref{fig:size}b), small particles display the behaviour reported above: the current is positive with a maximum around $f =0.75$ and turns off around $f = 1.5$. Sufficiently large particles (of diameter comparable to, but still less than, the ratchet wavelength) follow the same pattern with lower but still positive currents. When the size reaches one wavelength, the rectification is reduced to insignificant levels. 
Remarkably, however, intermediate size particles ($d\simeq 0.4$)  behave in a very different way: at large $f$, the current switches to negative values (Fig.~\ref{fig:size}b). This is despite the fact that for smaller particles, $\ell_D$ is too small for the system to show any significant rectification in passive mode.

\begin{figure}
\includegraphics[width = 0.9 \textwidth]{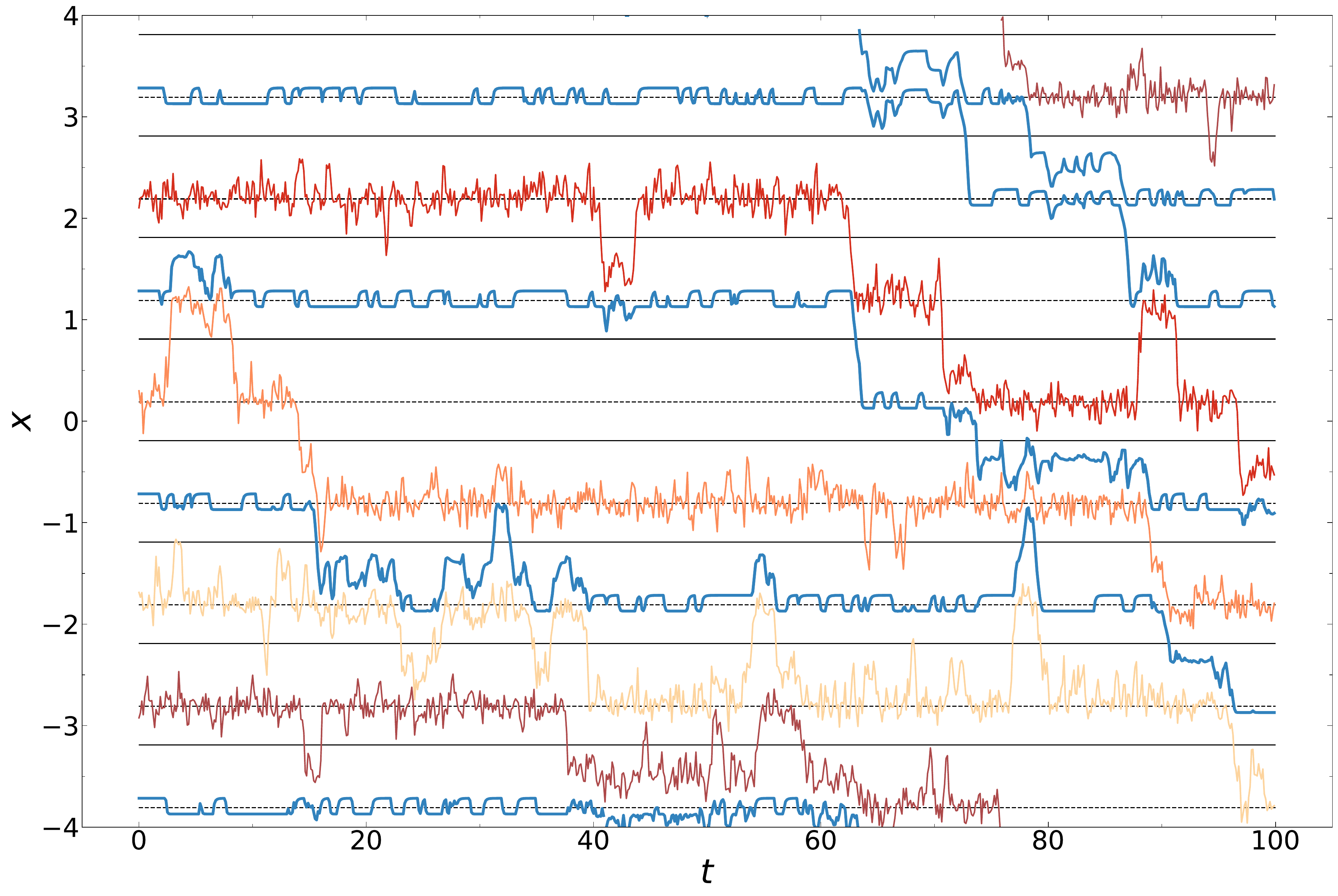}
\caption{Time series of 4 active (blue trajectories) and 4 passive (shades of orange and red) particles of intermediate size ($d = 0.4$) and with low passive diffusion coefficient ($\ell_D = 0.5$). The mean current is reversed for intermediate size particles. Dotted lines stand for the potential minima and the plain lines for the potential maxima.
Parameters: $f =2$, $\ell_D=0.5$, $\Gamma=1$, $\epsilon_{0} = 0.25$, $d = 0.4$, $N_a = N_p = 4$, $\phi_{a} = 0.5$, $c=1$, $t = 100$, $t_{\text{eq}} = 20$, $dt = 5 \times 10^{-6}$. }
\label{fig:seriet_d0.4_A0.5}
\end{figure}

 This phenomenon can be explained by considering the distance between the different maxima and minima in the potential landscape. The left maximum is located at a distance of $l_{l} \simeq 0.4$ from the bottom of a well while the right maximum is at $l_{r} \simeq 0.6$, as depicted in Fig.~\ref{fig:sawtooth_potential}. At high $f$, active particles stall in the well but passive particles have non-zero probabilities to jump to the left or right. When a passive particle does jump to the left, it will slide towards the minimum of its new well, and can push any active particle there through a distance of order $d$. This pushing can take the active particle across the left barrier, a process visible in Fig.~\ref{fig:seriet_d0.4_A0.5}. On the other hand, a passive particle jumping to the right can push an active particle in the well to the right, but not by far enough to cross the right barrier. Consequently, the leftward move is favored for intermediate size particles, enabling inversion of the current at smaller $\ell_D$ than is needed for smaller particles. When $\ell_D$ is raised so that the passive rectification model is anyway dominant, the size effect still contributes, so that intermediate size particles generate the largest passive-mode rectification as shown in Figs.~\ref{fig:size}c and \ref{fig:size}d.

\subsection{Role of quenched particle sequence}

As particles are not allowed to cross each other, the initial sequence of active and passive particles will be conserved dynamically. In the results reported so far, we averaged over random initial sequences. Here we address whether some arrangements could deliver stronger or weaker currents in the system. For instance, a sequence of several `cold' active particles in a row could hinder the mechanism for crossing a barrier when it involves the assistance of a hot passive particle. Due to exclusion, the resulting `cold clusters' might be capable of reducing the current globally.  In this Section, we check this point by studying the mean current for different sequences, addressing both the active and passive rectification modes. 

\begin{figure}
\includegraphics[width = 0.9 \textwidth]{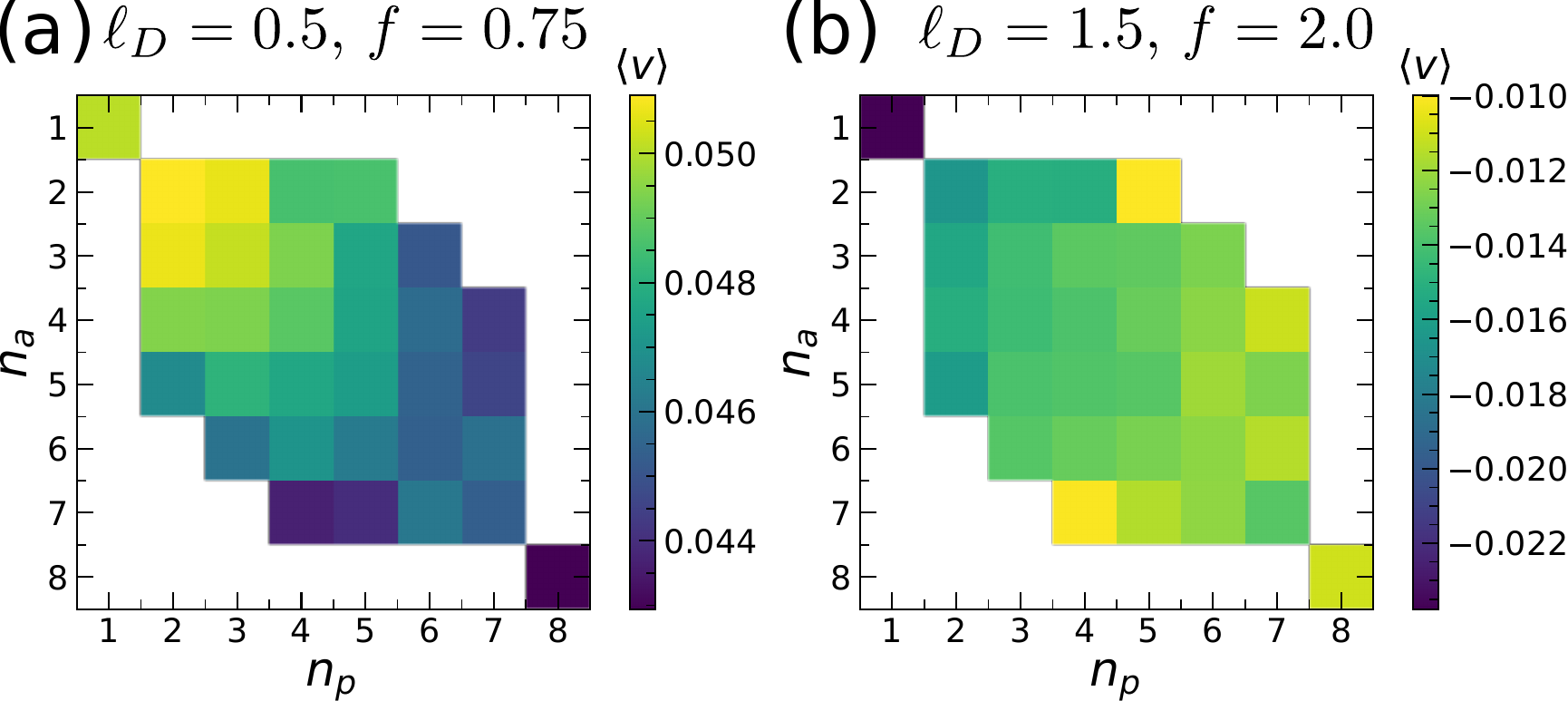}
\caption{Effect of sequence clustering on the current (a) for the active rectification ($\ell_D=0.5$, $f=0.75$) and (b) for the passive rectification ($\ell_D=1.5$, $f=2.0$).  
The quantities $\numAB,\numPB$ represent the numbers of consecutive active and passive particles in the system, see the main text for details.
Parameters: $\Gamma=1$, $\epsilon_{0} = 0.25$, $d = 0.2$, $N_a = N_p = 8$, $\phi_{a} = 0.5$, $c = 1$, $t = 400$, $t_{\text{eq}} = 100$, $dt = 5 \times 10^{-6}$, $N_{runs} = 26$ for every unique particles arrangement. }
\label{fig:clusters}
\end{figure}

Sequence space grows quickly with the number of particles in the system and its exploration becomes time-intensive. In order to reduce this intrinsic complexity, we consider the sequence of particle types within the system, and we extract the largest numbers of consecutive particles of each type, which we denote by $\numAB$, $\numPB$. (Consecutive particles are identified across periodic boundaries).

In Fig.~\ref{fig:clusters} we plot the average currents, binned according to the values of $\numAB,\numPB$, in a system of 16 particles (8 of each type).
For example, $(\numAB,\numPB)=(8,8)$ means that the active particles are arranged consecutively in the sequence while $(\numAB,\numPB)=(1,1)$ is the alternating arrangement. Each square in the figure was obtained by computing the average current over $N_{runs}$ trajectories of each unique sequence of particles, taking into account the translational invariance of the system. In a second step, results were grouped according to the largest active and passive clusters present in the sequence. This method ensures an equal statistical weight of each possible sequence in the computation of same-clusters current averages.

In the active rectification regime [$\ell_D=0.5$, $f=0.75$, Fig. \ref{fig:clusters}a], the differences between currents remain minor, varying by up to $\approx 10\%$. However, a trend is noticeable, in which the current drops slightly  in the presence of larger clusters. The diagonal symmetry of the plot indicates that this reduction of current results from a general crowding effect, which is not related to the particles' type (active or passive). This trend shows also that large clusters of passive particles do not impede strongly the current, as there is still a significant amount of rectification. A full active block is then able to push large clusters of passive particles in this system. 
When rectification is in passive mode, a similar trend can be observed. Currents are weaker in this mode, and their relative variation can reach 25 \% to 33\% (see Fig. \ref{fig:clusters}b). The case when all particles are alternated is particularly different from the other sequences, for which the current reaches a strong negative value. Even so, these variations from the typical currents are no more than a factor two.

The above results are for a relatively small system of particles ($N = 16$). Although clustering could have greater impact in larger systems, the probabilities of extreme cases (such as perfect alternation, or two monobloc clusters of active and passive particles) are themselves reduced exponentially.  Overall, the sequence dependence may merit further exploration, but we believe these results justify our decision to average over the initial sequence when presenting our main results in previous Sections.

\subsection{Comparison with passive-passive mixtures}\label{2temp}

In the previous Sections, we showed how the rectification mode depends separately on the diffusivities of active and passive particles via the tumbling rate and temperature respectively. Interestingly, a binary mixture of particles with different diffusivities can also be obtained by mixing passive particles at different temperature in the same ratchet potential.
Such mixtures can exhibit rich phenomenology, such as a phase separation \cite{grosberg_nonequilibrium_2015-1,ilker_phase_2020} reminiscent of active-passive mixture \cite{stenhammar_activity-induced_2015}. However, the analogy between AP and PP mixtures does not hold for every system or parameter ranges \cite{tanaka_hot_2017,grosberg_nonequilibrium_2015-1,ilker_phase_2020}. Hence we compare in depth both mixtures in the same ratchet potential.
 To avoid confusion, we will identify by the indices $1$ and $2$ the two sets of particles at different temperature in the passive-passive (PP) mixture, while reserving indices $a$ and $p$ for the different species in active-passive (AP) mixture. 

The PP mixture will be modelled as a mixture of Brownian particles with different diffusion coefficients.  
The dimensionless equations of motion are
\begin{align}
\frac{dx_{1,i}}{dt} & = F_{rat}(x_{1,i}) +  F_{int,i} + \ell_{D}^{(1)} \, \xi_i(t) \\
\frac{dx_{2,j}}{dt} & =  F_{rat}(x_{2,j})  +  F_{int,j} + \ell_{D}^{(2)} \, \xi_j(t) \,.
\end{align}
We have set the mobility ratio $\Gamma = 1$ as we did in the AP mixture, so the particles have equal mobility.
The effective temperatures of the two noise terms are in the ratio $\big[\ell_{D}^{(1)}/\ell_{D}^{(2)}\big]^2$.

To make a useful comparison between PP and AP systems, we need a method for choosing the parameters of the active particles and their passive replacements (species 1, say) to be `as similar as possible'. Clearly we can use the same densities, mixing ratios, and interaction parameters in which case we must decide how best to choose the passive diffusivity or temperature to match the active counterpart. 

Carrying out this matching in the ratchet system is complicated, as various forces are applied to particles: the ratchet force itself, the self-propulsion force for active particles, and the interaction forces. All those forces will impact the diffusion coefficients, while some will also cause rectification as we have seen. 
We can be guided by the noninteracting limit where the ratchet force affects drastically the diffusivity of active particles, suppressing it to zero when the maximum ratchet force exceeds the swim speed for both left and right movers.

To compare AP and PP mixtures effectively, it is then relevant to match single-particle diffusion coefficients in the external potential rather than in free space (as was done in Section~\ref{APM}). We introduce the diffusion coefficient of a single particle in the external potential $V(x)$ as 
\begin{equation}
D_{\spec}^V(t) = \frac{1}{2}\lim_{t \rightarrow \infty} \frac{d}{dt} \left( \langle y_{\spec}^2(t) \rangle - \langle y_{\spec}(t) \rangle^2\right)
\end{equation}
in which $y_{\alpha}$ stands for the position of a single particle only subjected to the ratchet force and the noise. 
For passive particles one has $\langle y_{\spec}(t) \rangle=0$, and the relation between bare and periodic-driven diffusivities is well established \cite{festa_diffusion_1978}:
\begin{equation}
\label{eq:driven_diff}
D_{p}^V = \frac{D_p}{\langle e^{V/k_BT} \rangle_\lambda \, \langle e^{-V/k_BT} \rangle_\lambda } 
\end{equation}
in which we introduced the spatial average over one spatial wavelength of the ratchet:
$ 
\langle g(x) \rangle_\lambda = \frac{1}{\lambda} \int_{0}^{\lambda} g(x) dx .
$ 

In contrast, computing the active diffusion coefficient for a periodic potential is a complicated analytical task \cite{doussal_velocity_2020}. Instead we compute $D_a^V$ by simulating a single active particle; then we fix $D_p$ for passive species 1 using \eqref{eq:driven_diff}, so that its $D_p^V$ matches $D_a^V$. This has to be done separately for each chosen value of the ratchet force $f$. When $f \geq 1.5$, the active particle is trapped in the potential well, and $D_a^V$ is zero. Active particles are then matched to passive particles at zero temperature ($\ell^{(1)}_D=0$) for those high values of $f$.
For the second species, we match $\ell^{(2)}_D$ to the passive species of the AP mixture, that is $\ell^{(2)}_D=\ell_D$. Numerical values of the different diffusion coefficients for each $f$ are reported in Table \ref{tab:matching_diff} of the Appendix.

\begin{figure}
\includegraphics[width = 0.8 \textwidth]{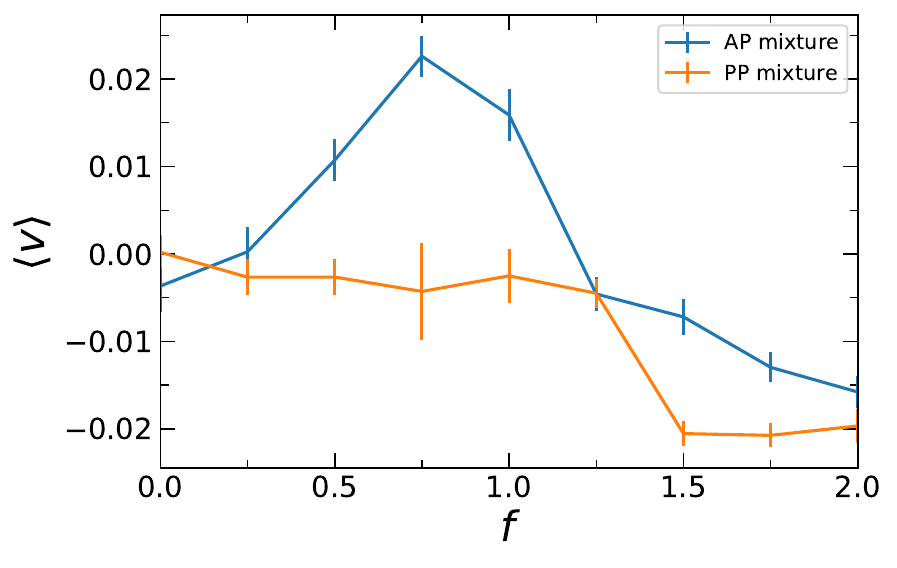}
\caption{Currents generated in AP and PP mixtures with matched values of the single particle diffusivity $D^V$ for each value of $f$. 
Parameters: $\Gamma=1$, $\epsilon_{0} = 0.25$, $d = 0.2$, $\ell_D = \ell_D^{(2)} = 1.5$, $N_a = N_p = N_1 = N_2 =4$, $\phi_{a} = \phi_{1} = 0.5$, $c = 1$, $t = 800$, $t_{\text{eq}} = 400$, $N_{runs} = 40$, $\alpha =1 $. The different values of $\ell_D^{(1)}$ used in the simulations of the PP mixture are reported in Table \ref{tab:matching_diff} (see Appendix). All simulations used a timestep $dt=5 \times 10^{-6}$ except for the PP mixture at $f=0.5$, $f=1$ ($dt=2.5 \times 10^{-7}$) and at $f=0.75$ ($dt=1.25 \times 10^{-7}$).}
\label{fig:AP-PP}
\end{figure}

Numerical results for the current as a function of the ratchet force, found by matching diffusivities in this manner, are shown in Fig.~\ref{fig:AP-PP} for the case of $\alpha = 1$. The main feature is that the active rectification mode is entirely missing for the PP mixture; the currents are negative across the entire range of $f$.  However, in the passive rectification regime at high $f$, hot passive particles push colder particles to the left -- and this happens whether the `cold' particles are active or passive. Hence, passive rectification is a more general phenomenon which works for a mixture of weakly diffusive and strongly diffusive particles, independently of their nature. Nonetheless, the negative current is stronger in the matched PP mixture than in the AP mixture.

These results support our earlier statement in Section \ref{APM} that there is no simple mapping between AP and PP cases.
Since the active rectification mode is always absent in PP mixtures, the only regime in which AP and PP can behave similarly across the full range of $f$ is the $\alpha \to\infty$ limit, where active rectification disappears even in the AP case. This limit corresponds to very `cold' active particles, which behave the same as passive particles with $\ell_D \rightarrow 0$. One has in effect a PP mixture with one species at negligible temperature. Away from this limit, there is no equivalence between AP and PP, even when the system is rectifying in the passive mode. 

In summary, active particles are central to the rectification process, either by generating positive currents in active rectification mode, or by resisting the pushing forces of passive particles in the passive rectification mode. Even in the latter, the effect of the active particles cannot be fully reproduced by replacing them with passive ones of the same single-particle diffusivity $D^V$ in the given potential $V(x)$.

\section{Conclusions}
\label{sec:conclusions}

In this paper, we studied the rectification of an active-passive mixture in a smooth ratchet potential, using a non-dimensionalised model of interacting RTPs and Brownian passive particles. The presence of excluding interactions, which prevent particle trajectories from crossing for both types of particle, generates richer phenomenology than in systems composed of active particles alone. 

First, the repulsive active particles can be used as a pumping device to generate a non-zero current of passive particles. This relies on the active rectification mode, which is absent for passive particles in thermal equilibrium, and which moves particles in the easy direction of the ratchet (the one with lower maximum barrier force). Second, the repulsive passive particles can be used to transport active particles (and themselves) in the opposite direction; this is not possible with a single species of active particle whatever its parameters. Although a passive mode is also seen for passive-passive mixtures at two different particle temperatures, the possibility of transiting from a forward active-dominated rectification to a reversed passive-dominated one, simply on varying the scale parameter $f$ for the ratchet force, is unique to the AP mixture. 

Mixing particles of different natures appears then as an interesting route to organizing their motion along a 1D channel. For example the active mode could be used to create a forward current of passive particles even when the latter are subjected to an additional backward force (creating a linear spatial ramp in their potential). This represents an interesting type of active engine which uses the bulk activity of the particles to do work everywhere, rather than only getting useful work from those particles that are near the bounding walls of the system \cite{ekeh_thermodynamic_2020}. The model introduced in this study helps to understand the role of different parameters in such a task. As established by numerical simulations, the current direction can be chosen by varying parameters. The most important ones can be viewed as are the diffusivities of the different species in the system, controlled through the quantities $\alpha$ and $\ell_D$. However, comparison with the PP-mixture with matched one-body diffusivities (whether matched in free space, or in the ratchet potential itself) confirmed that activity rather than just diffusivity is fundamental to the rectification problem, as for non-interacting active particles escaping a confining potential \cite{demaerel_active_2018,geiseler_kramers_2016}. The presence of an active force with finite persistence time is what makes the active rectification possible, but it also reduces the effectiveness of passive rectification by resisting the pushes of passive particles.

Our numerical simulations also revealed the role of particles size in the determining the current in the active-passive mixture. By choosing passive particles whose size is comparable to the distance between the bottom of a potential well and the top of the nearest barrier, it becomes possible to reverse the current into passive rectification mode in a regime where smaller particles show active rectification. Additional control of the current can be achieved selecting carefully the sequence of passive and active particles within the system, although highly atypical patterns (such as perfectly alternating or demixed into large blocks) appear necessary to achieve this. A perfectly alternating sequence maximises the reverse  current in passive rectification mode and to a less extent the forward current in active rectification mode. Longer consecutive sequences of active or passive particles tend to decrease the current amplitude in both modes.

The studies here do not represent a complete exploration of the high dimensional parameter space of our model. In the Appendix, we briefly address the effects of two further parameters, the active particle fraction $\phi_a$ and the energy parameter $\epsilon_0$ for the repulsive interaction. Neither has a strong qualitative effect, apart from an obvious downscaling of the active-mode current as $\phi_a$ is reduced. Since the main role of interactions is exclusion, and exclusion is present for any $\epsilon_0>0$, a strong effect of this parameter would be surprising.

For simplicity we have chosen certain parameters of the particles to be the same for the active and passive components. These include both size and mobility (defined as the ratio of particle velocity to net force). A natural extension of our work would be to make such parameters differ between species. Given the complicated interplay between diffusivity, activity, and size in determining the rectification currents reported above, making these parameters species-dependent could offer further routes to controlling the currents in the system. 

In our 1D model, the particles' arrangement is conserved for all time and active and passive currents are coupled through the single-file constraint. Extending this model by partially lifting the single-file constraint would be particularly interesting. This extension could be carried out by the introduction of a bounded interaction potential, or a non-zero crossing probability in the model. 
In such a system, we expect that active particles alone should still be rectified, leading to a positive current.  For a mixture, the crucial question is how effectively these active particles drag (or push) the passive ones across the  potential barriers.  This will likely depend on details of the interaction and the volume fraction.  To achieve negative currents, we recall that  ``cold'' active particles should be pushed over the potential barriers by ``hot'' passive ones.  It seems likely that such a process is still possible without any single-file constraint, but its dependence on model parameters is harder to predict.
We note that relaxation of the single-file constraint would have other consequences too.  For example, it enables new phenomena such as  demixing of active and passive particles, or clustering of the active ones, which are currently forbidden by the single-file constraint.

A related question concerns the extension of our work from 1D to higher dimensions.
 For asymmetric potentials such as  Eq.~\eqref{eq:sawtooth_pot} (supplemented by one additional dimension), there is no single-file constraint any more, so the arguments of the previous paragraph become relevant.  Previous work on 2D systems also suggest that currents may still flow in either direction.  For example, 
current reversal is observed~\cite{mcdermott_collective_2016} for active particles alone in a 2D potential, and passive currents were also seen in an AP mixture, in presence of disk-shaped obstacles \cite{rojas-vega_fast_2021}.
Of course, 2D systems also support more complex behaviour including dynamical clustering and phase separation~\cite{weber_binary_2016,stenhammar_activity-induced_2015}, which would also affect the ratchet currents.

Conversely, self-assembly and phase separation in 2D could also be strongly affected by introduction of external potentials as considered here. (This relates also to recent work on bulk active particles near a surface with periodic modulations \cite{nikola_active_2016}.)  Indeed, if one considers a series of ever narrower quasi-1D channels, at some point the physics reported here must be recovered. Our work therefore helps pave the way for better manipulation of AP mixtures, in various 1D and quasi-1D geometries, through the use of external potentials \cite{kumari_demixing_2017}.


\subsection*{Acknowledgments}
We thank Tirthankar Banerjee and Tal Agranov for useful discussions. This work was funded in part by the European Research Council (ERC) under the EU's Horizon 2020
Programme, Grant agreement No. 740269.
MEC is funded by the Royal Society.

\appendix
\section{}
\label{app}
\subsection{Diffusivity parameters for AP / PP comparison}
Table \ref{tab:matching_diff} shows the matching diffusivity parameters found using the protocol described in the main text. 

\begin{table}
\caption{\label{tab:matching_diff} Numerical values used to match the diffusivities of species 1 in the passive-passive mixture with the diffusivity of non-interacting active particles in the same ratchet potential. Other parameters are the same as mentioned in Fig. \ref{fig:AP-PP}. High ratchet force $f \geq 1.5$ leads to $D_a^V =0$, as active particles are trapped in the potential wells for those values of the force.}
\begin{ruledtabular}
\begin{tabular}{dddd}
\multicolumn{1}{c}{$f$} & \multicolumn{1}{c}{$D_a^V$(numerics)} & \multicolumn{1}{c}{Matching $\ell_D^{(1)}$} & \multicolumn{1}{c}{$\ell_D^{(2)}$} \\
\hline
0.00 & 5.15\times 10^{-1}& 1.02 & 1.50\\
0.25 & 4.80 \times 10^{-1}& 0.98& 1.50 \\
0.50 & 2.01\times 10^{0} & 2.00 & 1.50 \\
0.75 & 1.64 \times 10^{1} & 5.73 & 1.50 \\
1.00 & 5.27 \times 10^{-1} & 3.25 & 1.50 \\
1.25 & 8.35 \times 10^{-2}& 0.58 & 1.50 \\
\end{tabular}
\end{ruledtabular}
\end{table}

\subsection{Impact of $\epsilon_{0}$ on the currents}

\begin{figure}
\includegraphics[width = 0.9 \textwidth]{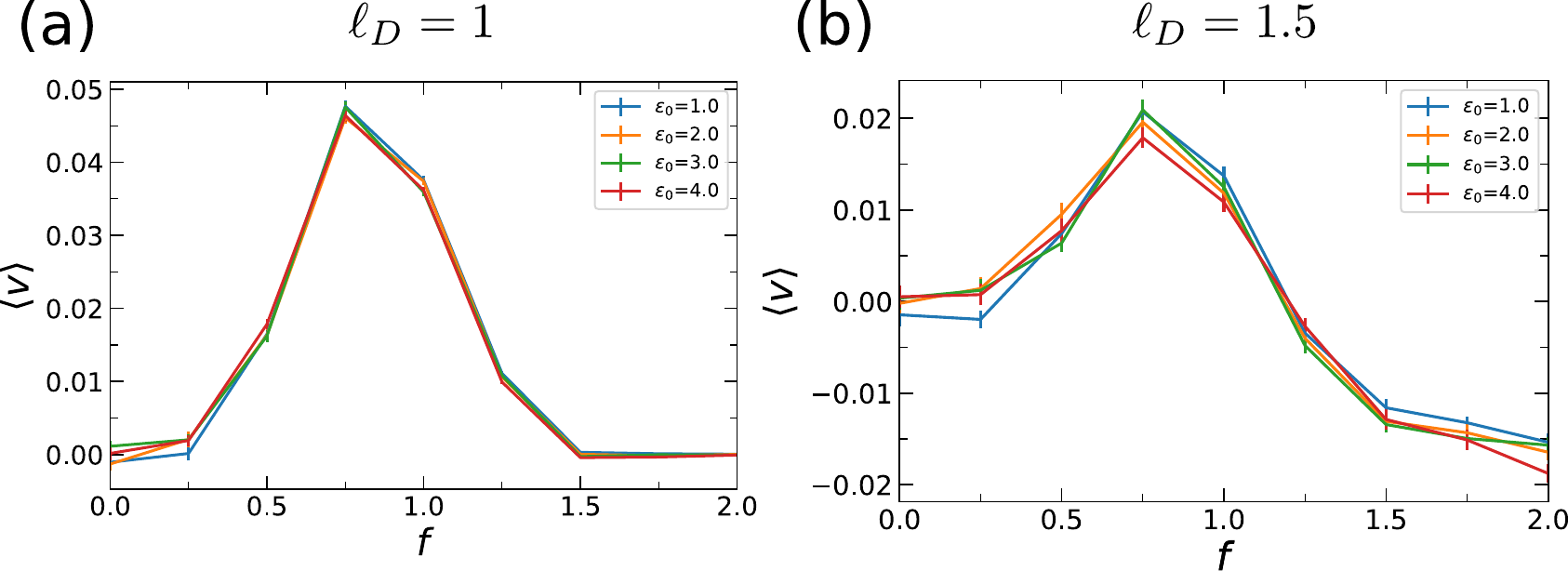}
\caption{Effect of interaction strength on the current as a function of ratchet force for two values of $\ell_D$: (a) $\ell_D = 1$, (b) $\ell_D = 1.5$. Parameters: $\Gamma=1$, $d = 0.2$, $N_a = N_p = 10$, $\phi_{a} = 0.5$, $c=1$, $t = 800$, $t_{\text{eq}} = 200$, $dt = 1.25 \times 10^{-6}$, $N_{runs} = 40$. }
\label{fig:interactions}
\end{figure}

The parameter $\epsilon_{0}$ measures the amplitude of interaction force in units of the self-propulsion force. This parameter does not impact strongly on the current curves as a function of $f$, as shown in Fig.~\ref{fig:interactions}. 

\subsection{Impact of $\phi_a$ on the currents}

\begin{figure}
\includegraphics[width = 0.9 \textwidth]{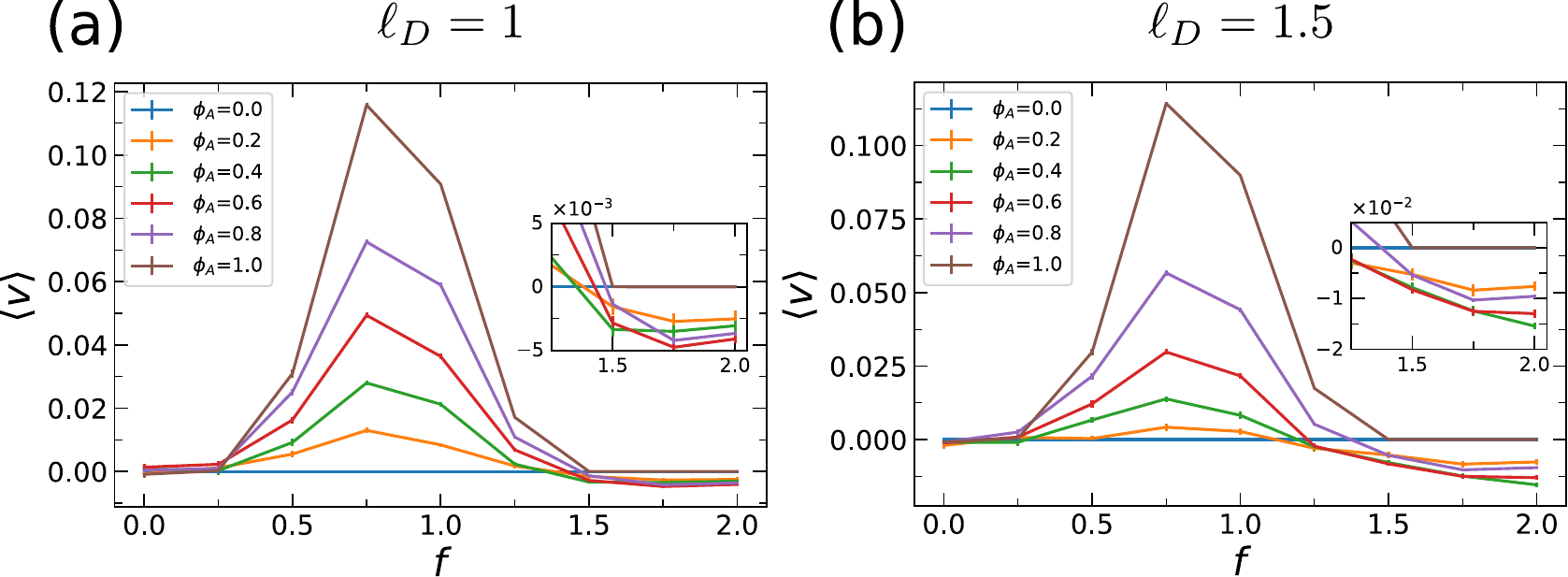}
\caption{Effect of active particle fraction ($\phi_a$) on the currents for two values of $\ell_D$: (a) $\ell_D = 1$, (b) $\ell_D = 1.5$.
Parameters: $\Gamma=1$, $\epsilon_{0} = 0.25$, $d = 0.25$, $N_{tot} = 20$, $c=1$, $t = 800$, $t_{\text{eq}} = 200$, $dt = 5 \times 10^{-6}$, $N_{runs} = 40$. }
\label{fig:wa}
\end{figure}

Here we briefly investigate by simulation the effect of active particle fraction $\phi_a$ on the current. As shown in Fig.~\ref{fig:wa}, the current grows non-linearly in the parameter range where active rectification occurs.
Similar observation was found for active and passive mixture in a pore, where currents vary non-linearly with the fraction of active particles \cite{ghosh_self-propelled_2013}. This trend in the pore is however different from the one found  here: in a pore, current can decrease with an higher fraction of active particles in the system, while in our setup, the active current per particle only increases with $\phi_a$.
Beyond the point where isolated active particles would become localised ($f \gtrsim 1.5$), one enters the passive rectification mode with negative currents. In this domain, the current is less affected by changing $\phi_a$ (see insets in Fig. \ref{fig:wa}).

\bibliographystyle{apsrev4-2}
\bibliography{Offline}

\end{document}